\documentclass[conference]{IEEEtran}
\usepackage{amsfonts}
\usepackage{amsmath}

\input epsf
\usepackage{graphicx}
\usepackage{siunitx}
\usepackage[ruled,vlined]{algorithm2e}
\usepackage[utf8]{inputenc}

\usepackage{threeparttable}

\begin{document}
\bibliographystyle{IEEEtran}

\title{\Huge A Reconfigurable Dual-Mode Tracking SAR ADC without Analog Subtraction}

\author{\authorblockN{Mehdi Safarpour\authorrefmark{1}, Reza Inanlou \authorrefmark{2}, Olli Silvén  \authorrefmark{1}, Timo Rahkonen\authorrefmark{1} and Omid Shoaei \authorrefmark{2}}
\authorblockA{\authorrefmark{1}Faculty of Electrical Engineering and Computer Science, University of Oulu, Oulu, Finland}\authorrefmark{2}School of Electrical and Computer Engineering, University of Tehran, Tehran, Iran}
\maketitle

\begin{abstract}
In this contribution, it is proposes to limit the quantization search space of a successive approximation analog-to-digital converter through an analytic derivation of maximum possible sample-to-sample variation. The presented example design of the proposed ADC is an 8-bit 1MS/s ADC with SAR logic customized to incorporate this priori information while no modification has been required to the analog circuitry. In comparison to conventional SAR conversion, the proposed tracking approach yields significant reduction in total power consumption in oversampling mode. The power savings are due to the reduced number of SAR cycles, and voltage variation minimization across DAC capacitors. The design is reconfigurable both to conventional SAR sampling and the proposed tracking scheme. The approach is attractive for SAR ADCs embedded in very low power micro-controllers.
\end{abstract}
\IEEEoverridecommandlockouts

\IEEEpeerreviewmaketitle

\section{Introduction}
Most Micro-Controller Units (MCU) incorporate an Analog-to-Digital Converter (ADC) \cite{instruments2007tms320x2833x}. Thanks to their high power efficiency and flexible sampling rate, Successive-Approximation-Register (SAR) ADCs are popular with low power MCUs, but their bit precision tends to be limited to 12 bits due to linearity issues. Consequently, the MCU vendors provide developers, ADCs with oversampling option to improve the resolution \cite{microelectronics2011rm0090}. In its simplest form, the resolution enhancement is achieved through  oversampling the signal by factor $M$ (over Nyquist-rate), followed by, e.g. Hogenauer structure and simple FIR filters \cite{fredenburg201290}. However, the flip-side can be a substantial increase of power dissipation  \cite{ding2018delta}. Moreover, in sensor readout interfaces, reconfigurable ADCs that provide different sampling and resolution options, are especially useful, where the ADC is time-multiplexed to digitize different types of input signal with different bandwidth and resolution requirements \cite{harpe2018hybrid}.  
\newline
 A modified oversampling scheme that reduces the power dissipation of the quantization process through cutting number of cycles for each conversion during oversampling, is proposed. The method is based on the observation that in oversampling mode, maximum sample-to-sample variation is constant and deterministic. In an attempt to take advantage of this observation, a tracking ADC is introduced to save power through constraining quantization levels that need to be resolved. The scheme can be an alternative for the commonly used oversampling techniques in conventional ADCs. Our design also supports regular Nyquist-rate sampling and can be reconfigured to various digitization schemes.
\newline
Previous designs generally attempt to reduce conversion cycles by quantizing only sample-to-sample variation which necessitate some form of analog subtraction, e.g. authors of \cite{lee2011input} propose a modification to the Digital-to-Analog-Convertor (DAC) in order to derive sample-to-sample variations, but this modification prevents employment of bottom plate sampling (and consequently facing charge injection problem).  The proposed method in this work uses prior knowledge to limit quantization search space, hence reducing number of cycles required to resolve each sample.  In this contribution, contrary to previous works, e.g. Noise Shaping SAR, the idea is not to use the ADC to digitize the difference between the current sample and previous converted value, but instead track the signal through adjust DAC value. In other words. the current conversion is held and is used to resolved the next sample, through minor modifications. 

\section{Theoretical Concept}
The purpose of the following is to derive an analytic expression for determining the maximum change between two consecutive samples, when the input signal has been low pass filtered and is oversampled by a factor of $M$. ADCs typically have a front-end anti-aliasing filter before quantization and it guarantees the input signal to be low-passed.  The fastest changing band-limited signal would be a sinusoid with maximum frequency and full dynamic range.
Let us assume the Nyquist-rate sampling frequency to be $f_s$ , oversampling factor $M$, and oversampling frequency $f_{os}$. When the highest frequency component in the signal is $f_{max}$, the sampling rate $f_s$ must be more than $2f_{max}$, but for our purpose in the calculations, one can safely be assume $f_s=2f_{max}$, leading to over-sampling rate of $f_{os}=2Mf_{max}$. The signal to be sampled can therefore be expressed as,
\begin{equation}
x(t)=Acos(2\pi f_{max} t+\phi)
\end{equation}
where $A$ is the amplitude, $t$ is time and $\phi$ is the phase. The oversampled signal is 
\begin{equation}
x[n]=Acos(  \frac {\pi n} {M}+\phi)
\end{equation}
and the difference between two adjacent samples is defined by,
\begin{equation}
D[n]=x[n]-x[n-1]
\end{equation}
To find the maximum value for the difference (3) in the oversampled signal, the first derivative of the difference is set equal to zero, then using backward differentiation one can solve for $n$:
\begin{equation}
\frac { \partial D}{\partial n} =\frac {(D[n]-D[n-1])}{1}
\end{equation}
\begin{equation}
\frac { \partial D}{\partial n} = Acos(\frac {\pi n} {M})-2Acos(\frac {\pi (n-1)} {M})+Acos(\frac {\pi (n-2)} {M})
\end{equation}
Setting the above equal to zero ($\frac { \partial D}{\partial n}=0$) gives,
\begin{equation}
 Acos(\frac {\pi n} {M} + \phi)-2Acos(\frac {\pi (n-1)} {M}+ \phi)+Acos(\frac {\pi (n-2)} {M}+ \phi)=0
\end{equation}
To simplify the notation, let $x=\frac{\pi n}{M}$ and $b = \frac{\pi}{M}$ , and using angle sum and difference identities, we get
\begin{equation}
\begin{split}Acos(x+\phi)-2Acos(x+\phi)cos(b) - 2Asin(x+\phi)(b) \\
+  Acos(x+\phi)(2b) + Asin(x+\phi)(2b) =0
  \end{split}
 \end{equation}
Let assume $M\gg \pi$ (which regularly is the case, in our experiments $M	> 16$), $b\approx 0$  and using Small Angle Approximation the following is resulted,
\begin{equation}
\begin{split}
Acos(x+\phi)-2Acos(x+\phi)(1-b^2/2)-2Asin(x+\phi)(b)\\ +Acos(x+\phi)(1-2b^2)+Asin(x+\phi)(2b)=0 
\end{split}
\end{equation}

Most of the terms in the above expression cancel out each other, leaving
\begin{equation}
-(b^2)Acos(x+\phi)=0 
\end{equation}
from which solving for $x$ yields
\begin{equation}
x=\frac{k \pi}{2} - \frac{M \phi}{\pi}, {k \in \mathbb{Z}}
\end{equation}
Therefore,  $D_{max}$  for $n=\frac{k \pi}{2} - \frac{M \phi}{\pi}$  becomes
\begin{equation}
D_{max}=\pm Asin(\pi/M)
\end{equation}
Small angle approximation can be applied again and hence the maximum range is approximated as $ \pm \frac{A\pi}{M}$. Having the current sample $x(nT_s)$  and its quantized equivalent, the next sample would be in the range limited to $\pm \frac{A\pi}{M}$,
\begin{equation}
x[(n+1) T_s ]=x[nT_s ]\pm \frac{A\pi}{M}  
\end{equation}
Hence, the search space of the ADC can safely be limited to the range of $\pm \frac{A\pi}{M}$. This can be exploited to reduce power consumption of ADCs embedded in MCUs. 
\section{Proposed ADC}
\subsection{General Description}

Figure 1 shows the conceptual diagram of the proposed ADC.  As shown, the main building blocks of the ADC consist of an 8-bit shift register, Z1 and Z2 delay generation unit, arithmetic unit and a successive approximation converter (ADC core). The ADC core itself is implemented by a binary weighted capacitor array that perform digital to analog conversion (DAC), a comparator and a passive sample-and-hold circuit.  The details and considerations of the unit capacitor in the DAC capacitor array as well as the comparator are similar to
 \cite{inanlou2013noise}; but the unit capacitor is 15fF in this design. Also, the arithmetic unit is composed of an 8-bit adder/subtract, an 8-bit delay cell and few digital logic gates for control and mode selection. Based on the mode pin status, the ADC operates in regular mode or oversampling mode. The functionality of the proposed ADC is described in the next subsection. 
\subsection{Sampling Modes}
As explained earlier, the proposed ADC digitizes the analog value in successive approximation manner, but instead of conventional logic, its range and steps are adjusted based on mode of operation. Algorithm~1 and Algorithm~2, present detailed procedural description of each mode. 

In regular SAR mode similar to conventional logic, the DAC value is set to zero at the end of each conversion but in tracking mode the DAC value is maintained. For regular SAR mode the user must set the initial value of the shift registers to “1000000” (for 8-bit ADC). For proposed tracking mode the initial value depends on the calculated maximum sample to sample variation:
\begin{figure}[bt]
\centering
\includegraphics[width=9.7cm]{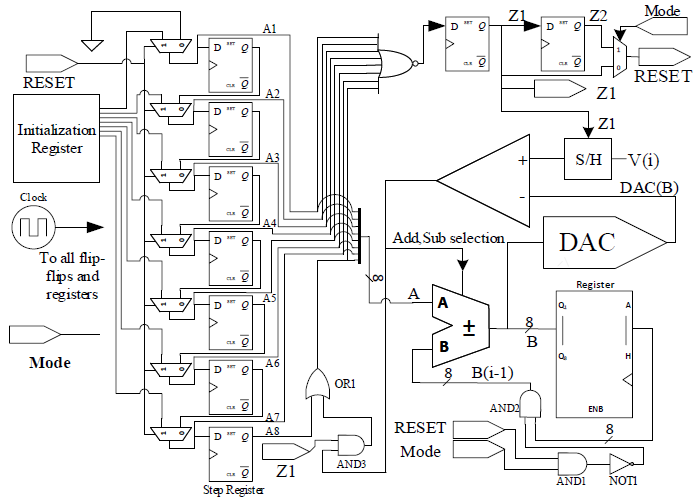}
\caption{Digital logic for realization of the proposed dual-mode SAR ADC.}
\end{figure}
\begin{equation}
\begin{split}
IntitialValue \geq \lceil \frac {D_{max} (Dynamic Range)} {(2^N-1)} \rceil \\ = \lceil \frac{2Asin(\frac{\pi}{M})}{2A}(2^N-1)  
\rceil 
  = \lceil sin( \frac{\pi}{M})(2^N-1) \rceil 
 \end{split}
 \end{equation}
Where $N$ is number of bits. For example, in case of an 8-bit ADC operating with oversampling rate of 64 the shift register initialization value is “00001000”. This means, in this example, only five cycles are required for conversion of each sample. The use of fewer cycles results in saving of energy required for each conversion. In addition, for most of code transitions the DAC does not need to be reset, which results in avoiding the discharge of MSB capacitors. This leads to extra power savings at DAC. Operation of the scheme is illustrated in Fig.~1 where the ADC is configured to function in regular SAR mode or in the proposed tracking SAR mode via  setting the Mode bit ‘1’ or ‘0’, respectively. In Fig.~1, Algorithm~1 and Algorithm~2, $B$ represents digital output, $B(i-1)$ denotes digital output from previous cycle, the Mode bit is set by user while RESET, Z1 and Z2 are generated within the circuit. 

The RESET signal is used to initialize the step shift register in both regular mode and in tracking mode, the signal from Z2 (multiplexed to RESET) is used to delay the initialization of the shift register by an additional clock cycle to let the comparator to compare the current analog sample with the quantized version of the previous sample. The design enables exploitation of sample to sample adjacency in oversampling mode, while maintaining the option to use the ADC also as a regular SAR. In both operation modes the delay flip-flop Z1 is used to indicate end of conversion.

\begin{algorithm}
\SetAlgoLined
\KwResult{Digital version, B, of the input analog sample}
Initialization Register (IR) is set by user; Mode is selected by user; B is binary output; DAC(B) is analog equivalent of B; $\textit{IR}$ $\leftarrow$ “10000000”
\newline  \textbf{Step 1.} Take sample, S/H $\leftarrow$  V(in) 
\newline  \textbf{Step 2.} StepRegister $\leftarrow$  $\textit{IR}$ 
\newline  \hspace{5cm} B(i-1)$ \leftarrow$ “00000000”
\newline  B $\leftarrow$ B(i-1)+$StepRegister$
\newline  \textbf{Step 3.}   \If{ StepRegister $\neq 0 $}{
   StepRegister $\leftarrow$ StepRegister $\gg 1$\;
    \eIf{ V(in)$>$DAC(B)}{
		B $ \leftarrow$ B(i-1)+ StepRegister\;   
        }
        {
		B $ \leftarrow$ B(i-1)-StepRegister\;   
  }
	jump to \textbf{Step.3}\;
   }
   \If{not(V(in)$>$DAC(B))}{
		B $ \leftarrow$ B(i-1)-1\;    
        }
 \caption{SAR ADC regular mode algorithm}
\end{algorithm}
In the regular mode the quantized value of a previous sample is ignored, while in the tracking mode the previous value on the DAC is held. In Fig. 1 blocks AND1, NOT1 and AND2 are used for this selection. The control signals are automatically generated using data in the shift register. At each clock cycle, data in the shift register is moved one bit to the right, generating a step value to be added or subtracted. An 8-bit asynchronous ripple-carry adder–subtracter is connected to the output of shift register (A1 being the MSB), and one clock cycle delayed version of quantized output, $B(i-1)$. The comparator’s output controls the addition/subtraction mode selection of the adder-subtracter. In addition, since in the last cycle all bits in the shift register have already been shifted out, Z1 is used as LSB (AND3 and OR1). Note, for applications that only require implementation of the tracking mode the circuit can significantly be simplified.

\begin{algorithm}
\SetAlgoLined
\KwResult{Digital version, B, of the input analog sample}
Initialization step is as described in Algorithm 1 but for tracking mode,  $IR \leftarrow \lceil sin(\pi /M)(2^N-1) \rceil$
\newline  \textbf{Step 1.} Take sample, S/H $\leftarrow$  V(in) 
\newline   StepRegister $\leftarrow$  $\textit{IR}$ 
\newline (note that the DAC value and the output, B, are held from previous conversion)
\newline  B $\leftarrow$ B(i-1)+$StepRegister$
\newline  \textbf{Step 2.}   \If{ StepRegister $\neq 0 $}{
    \eIf{ V(in)$>$DAC(B)}{
		B $ \leftarrow$ B(i-1)+ StepRegister\;   
        }
        {
		B $ \leftarrow$ B(i-1)-StepRegister\;   
  }
	jump to \textbf{Step.2}\;
   }
   \If{not(V(in)$>$DAC(B))}{
		B $ \leftarrow$ B(i-1)-1\;    
        }
 \caption{Tracking mode algorithm}
\end{algorithm}

\section{Results and Discussion}
Initially, a behavioral simulation using MATLAB was carried out to verify the functionality of the proposed ADC. Subsequently, a transistor level circuit simulation in 90~nm CMOS process with an HSPICE model was carried out to investigate the power consumption in both regular and proposed oversampling mode with different oversampling ratios. Specifications of the design is presented in Table I. The power consumption for different oversampling ratios were estimated through simulations. The summaries of the findings are in Table~\ref{table2} which shows energy consumption per sample reduces proportionally to the oversampling ratio. Figure~2 shows the power spectral density of the proposed ADC architecture in tracking mode for $OSR=64$.


\begin{table}[bt]
\caption{Properties summary of the sample SAR ADC.}
\begin{center}
\begin{tabular}{lc}
\hline
{Specification} & {Value }\\
Technology & 90nm CMOS \\
Resolution & 8  \\
ENOB* & 7.4  \\
Supply Voltage & 1V  \\
Max. Sampling Rate & 1 MS/s \\
Power Consumption & 12.9 $\mathrm{\mu}$W \\
\hline  
\scriptsize {*Regualar mode, Nyquist rate.}
\end{tabular}
\end{center}
\end{table}

The proposed architecture serves as an example on exploiting sample-to-sample variation without having to carry out analog subtraction. Although similar ideas have been investigated before \cite{lee2011input}\cite{lee2015power}, to the best knowledge of authors, general approach of the previous works was to quantize the analog difference of two consecutive samples while in this work the search space is adjusted digitally. Analog subtraction requires extra analog circuitry, which introduces more design constraints. In addition, the proposed approach can be reconfigured to conventional SAR logic mode (either for Nyquist-rate sampling or oversampling). Furthermore, though the digital complexity of the design is slightly higher that conventional SAR logic, the power saving can compensate this to a large degree. Moreover, low power digital design techniques can be employed without having to compromise the performance, to further reduce power consumption.
\newline
Although, the presented implementation of the design does not achieve the best FoM in Table~\ref{table3}, where comparison with similar works including Noise-Shaping (NS) SAR ADCs, is provided. However, as indicated before, we aimed at a reconfigurable architecture that can function as a regular SAR with more efficient oversampling mode via taking advantage of limited sampling-to-sample change. Please note, as it is presented in Table~\ref{table2}, in the proposed design, the power dissipation per sample scales down with increasing the sampling rate. Furthermore due to large logic section, further power lessening through voltage scaling can be  envisioned.
\begin{figure}[bt]
\centering
\includegraphics[width=9.5cm]{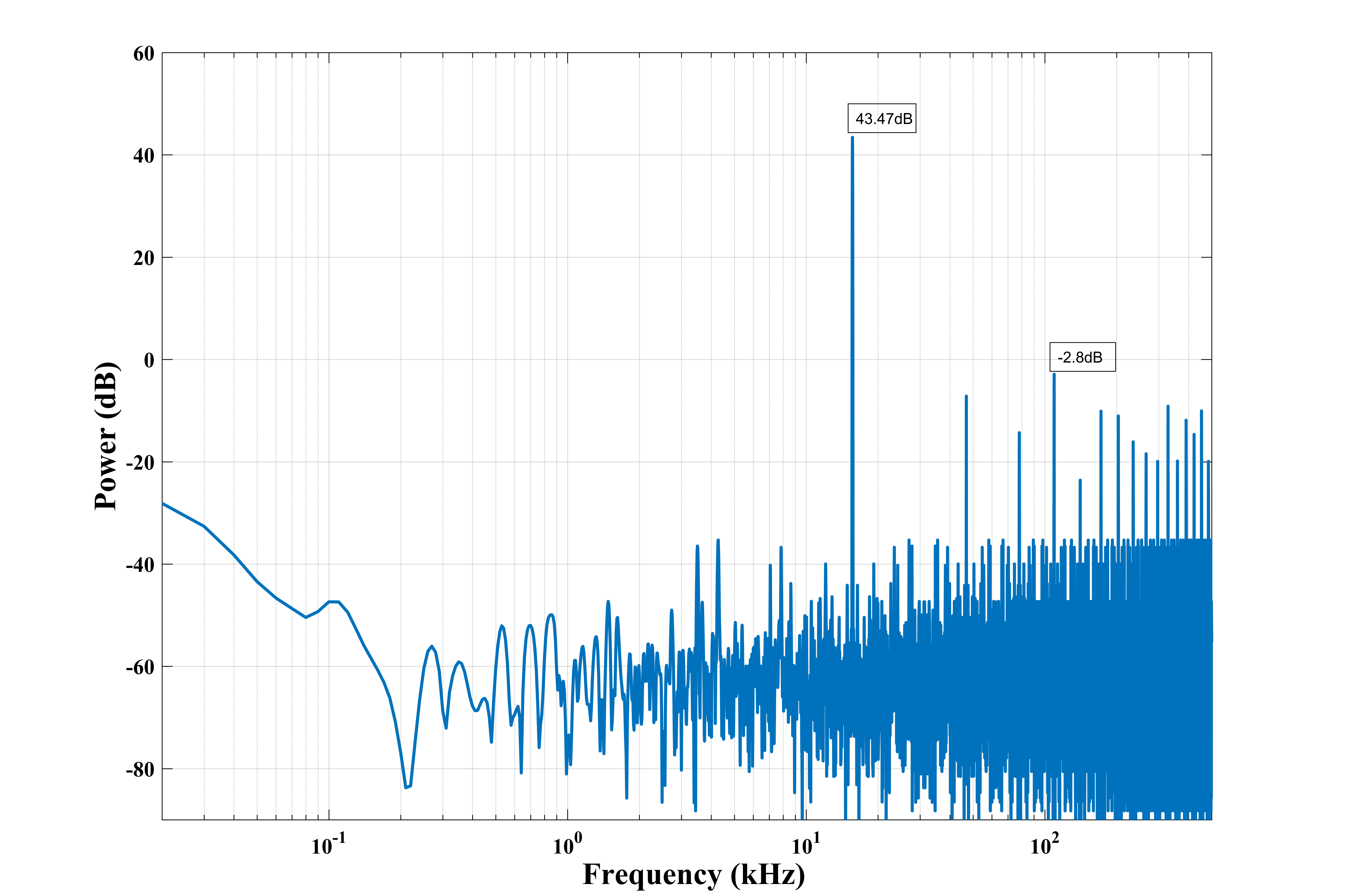}
\caption{Power Spectrum of the ADC in proposed oversampling mode with oversampling ratio of 64 (SFDR=46.3dB)}
\end{figure}

\begin{table}[bt]
\caption{Estimated Power Consumption With Respect To OSRs}
\begin{center}
\label{table2}
\begin{tabular}{ | m{1.1cm} | m{0.8cm}| m{1cm}| m{1.2cm}  | m{0.7cm} | m{1.1cm} | } 
\hline
 {OS Mode} & OSR* & Variation& Initial & Cycles & pW/S  \\ 
\hline
 {Regular} & Any & Max. & 10000000 & 9 & 12.8 \\
\hline
 {Tracking} & 32 & 25 & 00100000 & 7& 8.03 \\
\hline
 {Tracking} & 64 & 12.5 & 00001000 & 5& 6.33 \\
\hline
 {Tracking} & 246 & 3.1 & 00000100 & 4& 5.53 \\\hline
\hline
\multicolumn{6}{l}{$^{\mathrm{*}}$ \scriptsize {master clock was kept constant and the input frequency was scaled down.}}
\end{tabular}
\end{center}
\end{table}

\begin{table}[]
\caption{Comparison with the related works}
\centering
\label{table3}
\begin{tabular}{l|c|c|c|c|}
\cline{2-5}
          & \begin{tabular}[c]{@{}c@{}}Ref. \cite{Tong2019} \\ (Sim.)\end{tabular} &  \begin{tabular}[c]{@{}c@{}}Ref. \cite{shen2018reconfigurable} \\ (Chip)\end{tabular} & \begin{tabular}[c]{@{}c@{}}Ref. \cite{fredenburg201290} \\ (Chip)\end{tabular} & \begin{tabular}[c]{@{}c@{}}This Work\\ (Sim.)\end{tabular} \\ \hline
\multicolumn{1}{|l|}{Bandwidth }   & 60 KHz  & 20 MHz & 11 MHz  & 2-500 KHz                                                       \\ \hline
\multicolumn{1}{|l|}{ENOB}   & 9.55 & 9.8  & 10   & 7.4                                                        \\ \hline
\multicolumn{1}{|l|}{Power} & 2.97 $\mu$W & 1.8$m$W & 806 $\mu$W & 13 $\mu$W                \\ \hline
\multicolumn{1}{|l|}{Technology}   & 180~nm  & 180~nm & 65~nm & 90~nm   \\ \hline
\multicolumn{1}{|l|}{FoM}   & 36.9~f  & 70.2~f & 35.8~f   & 76.9~f    \\ \hline
\multicolumn{1}{|l|}{Architecture} & SAR   & Reconfigurable    & NS-SAR & Tracking SAR                               \\ \hline
\multicolumn{1}{|l|}{FOM}          & \multicolumn{4}{c|}{ $\frac{Power}{B.W \times 2^{ENOB} } $}    
\\ \hline
\end{tabular}
\end{table}

\section{Conclusion}
Taking advantage of how sample-to-sample variation is limited can provide for substantial energy efficiency improvements for SAR ADCs. Our contribution shows how this can be accomplished by modifying only the digital parts of the conversion logic, and minimum restriction is imposed on analog circuits. The digital section can enjoy power reduction techniques such as voltage scaling without having to compromise circuit functionality.  The presented digital logic supports both regular SAR and proposed tracking mode of operations. The method is useful in multi-channel ADCs used in embedded micro-controllers.

\section*{Acknowledgment}
The support of Academy of Finland for project ICONICAL (grant 313467) and 6Genesis Flagship (grant 318927) is gratefully acknowledged. 

\bibliographystyle{unsrt}

\bibliography{bib.bib}

\begin{thebibliography}{10}
\providecommand{\url}[1]{#1}
\csname url@samestyle\endcsname
\providecommand{\newblock}{\relax}
\providecommand{\bibinfo}[2]{#2}
\providecommand{\BIBentrySTDinterwordspacing}{\spaceskip=0pt\relax}
\providecommand{\BIBentryALTinterwordstretchfactor}{4}
\providecommand{\BIBentryALTinterwordspacing}{\spaceskip=\fontdimen2\font plus
\BIBentryALTinterwordstretchfactor\fontdimen3\font minus
  \fontdimen4\font\relax}
\providecommand{\BIBforeignlanguage}[2]{{%
\expandafter\ifx\csname l@#1\endcsname\relax
\typeout{** WARNING: IEEEtran.bst: No hyphenation pattern has been}%
\typeout{** loaded for the language `#1'. Using the pattern for}%
\typeout{** the default language instead.}%
\else
\language=\csname l@#1\endcsname
\fi
#2}}
\providecommand{\BIBdecl}{\relax}
\BIBdecl

\bibitem{instruments2007tms320x2833x}
{Texas Instruments}, ``{TMS320x2833x analog-to-digital converter (ADC) module
  reference guide},'' \emph{Literature Number: SPRU812A}, 2007.

\bibitem{microelectronics2011rm0090}
{STMicroelectronics}, ``{RM0090 Reference manual STM32F405xx, STM32F407xx,
  STM32F415xx and STM32F417xx advanced ARM-based 32-bit MCUs},'' September
  2011.

\bibitem{fredenburg201290}
J.~A. Fredenburg and M.~P. Flynn, ``A 90-ms/s 11-mhz-bandwidth 62-db sndr
  noise-shaping sar adc,'' \emph{IEEE Journal of Solid-State Circuits},
  vol.~47, no.~12, pp. 2898--2904, 2012.

\bibitem{ding2018delta}
Z.~Ding, X.~Zhou, and Q.~Li, ``{Delta-Measurement Low-Power SAR ADC
  Architecture with Adaptive Threshold-First Switching},'' in \emph{2018 IEEE
  International Symposium on Circuits and Systems (ISCAS)}.\hskip 1em plus
  0.5em minus 0.4em\relax IEEE, 2018, pp. 1--4.

\bibitem{harpe2018hybrid}
P.~Harpe, K.~A. Makinwa, and A.~Baschirotto, \emph{Hybrid {ADCs}, Smart Sensors
  for the IoT, and Sub-1V \& Advanced Node Analog Circuit Design}.\hskip 1em
  plus 0.5em minus 0.4em\relax Springer, 2018.

\bibitem{lee2011input}
B.-G. Lee and S.-G. Lee, ``{Input-tracking DAC for low-power high-linearity SAR
  ADC},'' \emph{Electronics letters}, vol.~47, no.~16, pp. 911--913, 2011.

\bibitem{inanlou2013noise}
R.~Inanlou, M.~Shahghasemi, and M.~Yavari, ``{A noise-shaping SAR ADC for
  energy limited applications in 90 nm CMOS technology},'' \emph{Analog
  Integrated Circuits and Signal Processing}, vol.~77, no.~2, pp. 257--269,
  2013.

\bibitem{lee2015power}
B.-G. Lee, ``{Power and Bandwidth Scalable 10-b 30-MS/s SAR ADC},'' \emph{IEEE
  Transactions on Very Large Scale Integration (VLSI) Systems}, vol.~23, no.~6,
  pp. 1103--1110, 2015.

\bibitem{Tong2019}
X.~Tong, M.~Song, Y.~Chen, and S.~Dong, ``{A 10-Bit 120 kS/s SAR ADC Without
  Reset Energy for Biomedical Electronics},'' \emph{Circuits, Systems, and
  Signal Processing}, May 2019.

\bibitem{shen2018reconfigurable}
Y.~Shen, Z.~Zhu, S.~Liu, and Y.~Yang, ``A reconfigurable 10-to-12-b
  80-to-20-ms/s bandwidth scalable sar adc,'' \emph{IEEE Transactions on
  Circuits and Systems I: Regular Papers}, vol.~65, no.~1, pp. 51--60, 2018.

\end{thebibliography}

\smallskip

\end{document}